\documentclass[aps,prl,twocolumn,showpacs,amsmath,amssymb,letterpaper]{revtex4-1}

\usepackage[pdftex]{graphicx} 
\usepackage{dcolumn} 
\usepackage{bm} 
\usepackage{relsize}
\usepackage{amsmath}
\usepackage{color}
\usepackage{psfrag}
\usepackage{epstopdf}
\usepackage{tocloft}
\usepackage{wasysym}
\definecolor{olive}{rgb}{0,0.6,0.4}

\begin{document}

\title{Inelastic electron tunneling spectroscopy of local ``spin accumulation'' devices}
\author{Holly N. Tinkey}
\affiliation{Dept. of Physics and CNAM, U. of Maryland, College Park, Maryland, 20742}
\author{Pengke Li}
\affiliation{Dept. of Physics and CNAM, U. of Maryland, College Park, Maryland, 20742}
\author{Ian Appelbaum}
\email{appelbaum@physics.umd.edu}
\affiliation{Dept. of Physics and CNAM, U. of Maryland, College Park, Maryland, 20742}

\begin{abstract}
We investigate the origin of purported ``spin accumulation'' signals observed in local ``three-terminal'' (3T) measurements of ferromagnet/insulator/n-Si tunnel junctions using inelastic electron tunneling spectroscopy (IETS). Voltage bias and magnetic field dependences of the IET spectra were found to account for the dominant contribution to 3T magnetoresistance signals, thus indicating that it arises from inelastic tunneling through impurities and defects at junction interfaces and within the barrier, rather than from spin accumulation due to pure elastic tunneling into bulk Si as has been previously assumed.
\end{abstract}
\maketitle

Creating, controlling, and detecting spin-polarized electron currents in nonmagnetic materials is the first step toward integrating spintronic devices with new functionalities and energy efficiency \cite{Zutic_RMP2004, Fabian_APS2007} such as spin transistors \cite{Datta_APL1990} and spin-based logic circuits \cite{Behin_nnano2010}, which make use of the electron's spin degree of freedom instead of its charge. Silicon (Si) has low spin-orbit coupling and long spin lifetime \cite{Huang_PRL2007}, making it an excellent candidate for spin-enabled devices, but the conductivity mismatch between semiconductors and ferromagnetic metal (FM) spin sources prohibits ohmic electrical spin injection and detection in this material \cite{Schmidt_PRB2000}. To overcome this difficulty, ballistic hot electron injection and detection techniques were used to finally achieve long-distance spin transport and coherent precession in intrinsic Si in 2007 \cite{Appelbaum_Nature07, Huang_PRL2007}. Using a tunneling barrier approach \cite{Rashba_PRB2000}, four-terminal nonlocal measurements of open-circuit voltage at ferromagnetic contacts\cite{Johnson_PRL1985} were subsequently demonstrated in degenerately doped Si at low temperature \cite{Erve_APL2007} and at room temperature in 2011. \cite{Suzuki_APE2011}

\begin{figure}[t!]
\begin{center}
\includegraphics[width=8.75cm, height=11.5cm]{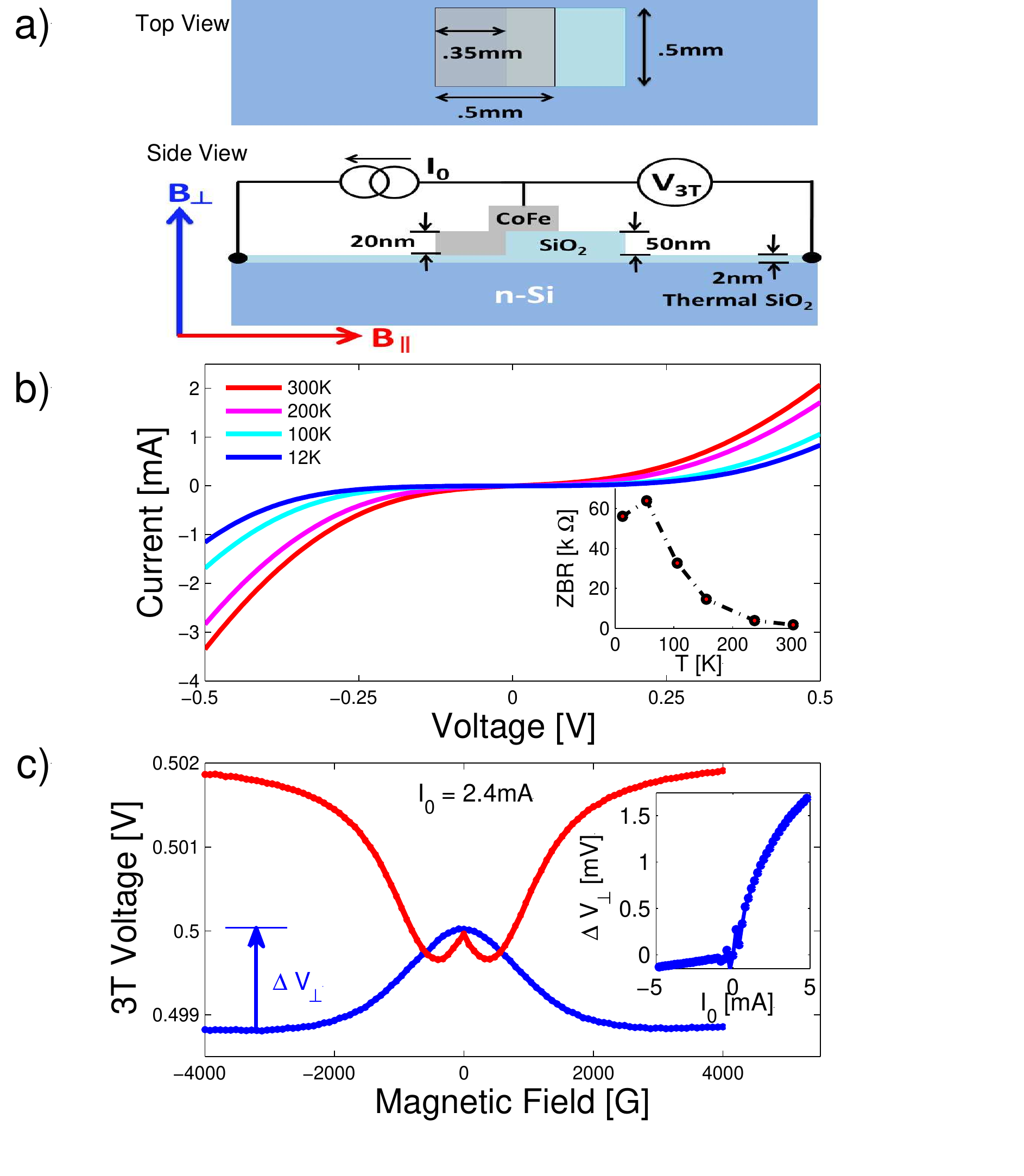}
\vspace{-20pt}
\caption{(Color online) (a) Top-down and side view of a junction and schematic of the 3T measurement. (b) Current-voltage curves and zero bias resistance (ZBR, inset) of a CoFe/SiO$_2$/n-Si junction at different temperatures. (c) 3T voltage signals in perpendicular (blue) and parallel (red) magnetic field at room temperature with 2.4mA applied current. Inset: 3T signal magnitude in perpendicular field, $\Delta V_\perp$, as a function of applied current. All data is from Contact 13.\label{FIG1}} 
\vspace{-25pt}
\end{center}
\end{figure}

During this time, there were also several claims that ``accumulation'' of spin-polarized electron density had been measured in highly doped Si \cite{Dash_Nature2009,Li_ncomm2011} using a local, three terminal (3T) geometry (schematically shown in Fig. \ref{FIG1}(a)) at and beyond room temperature. This setup is intended to employ the same ferromagnetic contact for simultaneous injection and detection with signals dependent on magnetic field-induced spin precession and dephasing (``Hanle effect''). These reports present magnetoresistance (MR) in kOersted magnetic fields on the order of 0.1\% (typically millivolt changes on one volt background at milliAmp constant current \cite{Dash_Nature2009,Li_ncomm2011,Jansen_PRB2010,Jeon_APL2011}), far exceeding voltage signals from non-local experiments \cite{Suzuki_APE2011}. The spin lifetimes ($\tau_s$) extracted from their magnetic field linewidths are two orders of magnitude lower than those measured by electron spin resonance (ESR) \cite{Pifer_PRB1975,Zarafis_PRB1998} under similar conditions and show little dependence on the carrier type \cite{Gray_APL2011}, the doping of the semiconductor in the transport channel, or temperature \cite{Dash_Nature2009, Li_ncomm2011}; the signal width also remains invariant in metals with considerably different spin-orbit interaction strength \cite{Txoperena_APL2013}. Additionally, the claimed lifetime and signal magnitude are not self-consistent; a simple theoretical model of elastic tunneling into the Si conduction band constrains signals $\apprle\frac{J\tau_s}{q^2n\mathcal{L}}\mathfrak{E}$, where $J$ is current density, $n$ is free carrier density, $q$ is the fundamental charge, $\mathcal{L}$ is the transport depth, and $\mathfrak{E}$ is the relevant energy scale (thermal energy for nondegenerate conditions, Fermi energy for degenerate). Typical parameters give expected 3T signal values in the microvolt regime, orders of magnitude smaller than measured.\cite{Dash_Nature2009,Li_ncomm2011,Jansen_PRB2010,Jeon_APL2011,Gray_APL2011,Txoperena_APL2013}

The fundamental difference between this local technique and the non-local measurements is that the detector junction in the 3T geometry is voltage-biased and substantial current is tunneling through it, making transport measurements susceptible to \emph{inelastic} tunneling contributions from impurities and defects in the barrier.

To quantify such contributions from inelastic tunneling to the 3T magnetoresistance, we employ inelastic electron tunneling spectroscopy (IETS), an experimental technique used to measure low-lying excitations in a variety of tunneling systems including p-n junction tunnel diodes, superconducting films, and metal-insulator-metal tunnel junctions \cite{Adkins_JPhysC1985, Brown_JChem1979}. At electrostatic potential energies provided by voltage bias corresponding to specific resonances, secondary inelastic tunneling pathways will open up, increasing the total conductance of the junction. This manifests itself most clearly as a thermally-broadened peak in the second derivative of the junction's $I(V)$ curve, measured in practice by applying a modulated current or voltage across the junction and detecting the second harmonic response with a lock-in amplifier. 

The devices used in this study were prepared on chips cleaved from n(arsenic)-doped Si wafers with resistivity .001-.005 $\Omega\cdot$cm and electron concentration $~$2$\cdot10^{19} \text{cm}^{-3}$. After etching with dilute HF, samples were placed on a hotplate in air at $480^{\circ}$C for 8 minutes to thermally oxidize a tunneling SiO$_2$ barrier on the surface. To fabricate ferromagnetic contacts, a 50 nm Si$\text{O}_2$ layer and 20 nm cobalt-iron ($\text{Co}_{88}\text{Fe}_{12}$) layer were deposited at different angles in UHV via electron beam thermal evaporation (deposition pressure$~$1-4 $\cdot 10^{-8}$ mbar) with a $500\times 500\mu$m$^2$ shadow mask array pattern such that the layers were shifted laterally as shown in the top and side views of Fig. \ref{FIG1}(a); the thick Si$\text{O}_2$ layer protected and electrically isolated the tunneling oxide layer during wirebonding. Before any measurements were taken, two devices on each sample were shorted to the substrate by driving an increasing current between them until dielectric breakdown of the thermal Si$\text{O}_2$ yielded ohmic contacts to eliminate in-series voltage drops from the junction resistance measurement. Fig. \ref{FIG1}(b) shows characteristic IV curves and zero bias resistance (ZBR) at different temperatures; for this junction in particular, the ZBR increases with decreasing temperature from 1.5k$\Omega$ at RT to over 60k$\Omega$ at 10K, which is consistent with tunneling into a barely non-degenerate semiconductor.

Measurements were taken in a 3T, local geometry shown in Fig. \ref{FIG1}(a) as a function of magnetic field at room temperature. Constant current ($I_0$) was injected through the shorted contact such that positive (negative) current and bias indicates electron injection (extraction) into Si through the SiO$_2$. As shown in Fig. \ref{FIG1}(c), under magnetic fields normal to the sample surface and bias conditions corresponding to electron injection from the FM contact, the 3T voltage took a quasi-Lorentzian shape. If fit to the spin accumulation model $\Delta V(B_\perp)=\Delta V_\perp$/$(1+(\omega_L\tau)^2)$ (Fourier transform of exponential spin decay, where $\omega_L=g\mu_BB/\hbar$ is Larmor frequency\cite{Jang_PRL2009, Huang_PRB2010}) the lifetime $\tau$ ranges from 50-70ps, far lower than values reported for ESR measurements of heavily As-doped Si \cite{Pifer_PRB1975}.  As shown in the inset to Fig.\ref{FIG1}(c), a significant $\Delta V_\perp$ appears under positive $I_0$ (electron injection), but does not exhibit the linear response expected from elastic tunnel injection\cite{FertIEEE2007} and is severely reduced for electron extraction. When the field is applied in-plane, the 3T signal inverts, with an unusual $\propto |B_{||}|$ linear regime at low fields at room temperature that becomes quadratic at lower temperatures (not shown) \cite{Hu_NatureMat2008}.

\begin{figure}[t!]
\begin{center}
\includegraphics[width=7.5cm, height=5cm]{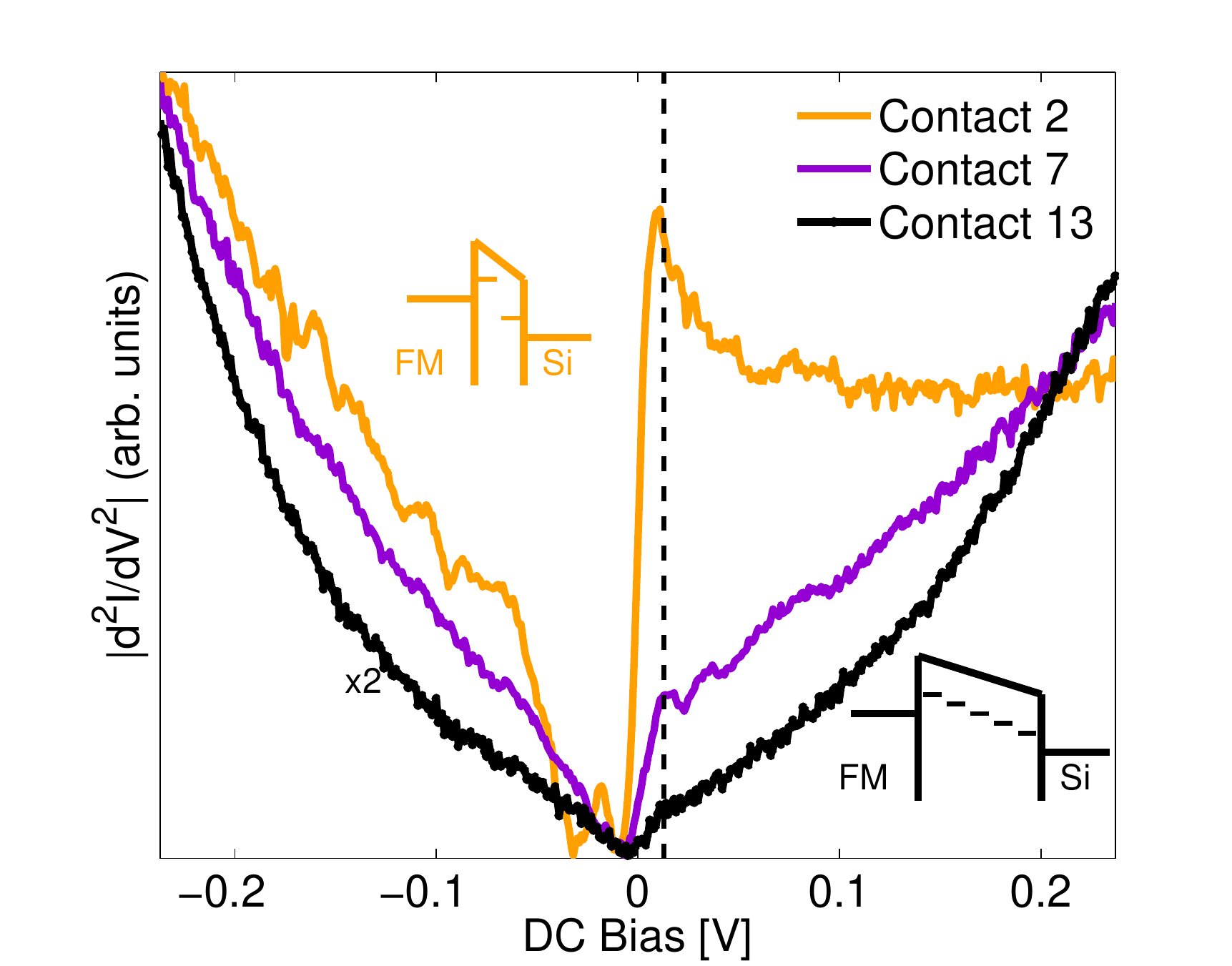}
\caption{(Color online) IET spectra for three contacts with various resistances on the same sample chip. Measurements taken at 15K, $V_{AC}$=2mV, $f_{AC}$=931Hz.  The dashed line at 13mV highlights a common feature in all spectra. Insets show schematic distribution of impurity levels in corresponding thick and thin barriers, modifying IETS structure.\label{FIG2}} 
\vspace{-20pt}
\end{center}
\end{figure}

Since a large rise in ZBR at low temperatures alone does not conclusively prove the existence of secondary tunneling states, IETS is used for more detailed analysis of the $I(V)$ structure. A DC+AC voltage of fixed modulation frequency was applied across the tunnel junction, and the corresponding second harmonic response from the junction was detected with a lock-in amplifier. The low-temperature (15~K) IET spectra of several tunnel junctions of varying resistance are shown in Fig. \ref{FIG2}. There exists some correspondence between the structure in the spectra of Contact 2 and 7 and known phonon modes in Si (46.3, 53.4 and 65.3 mV) and Si$\text{O}_2$ (138.4, 147.0, 153.5, 160.8, and 170.9mV) \cite{Salace_JAP2002}. However, besides a small peak that appears at low positive bias near 13mV, there are no common features in these spectra to identify a specific impurity or vibrational mode, despite an intrinsic measurement resolution of 5.4$k_BT\approx$5mV \cite{Adkins_JPhysC1985}. In general, devices with lower ZBR display smaller $\Delta V_\perp/V$ and higher intensity IET spectra with more distinct features, whereas more resistive devices have higher $\Delta V_\perp/V$ and smoother IET spectra. For comparison, at 12K contacts 2, 7, and 13 have ZBR of 950$\Omega$, 5.5k$\Omega$ and 55k$\Omega$ and showed $\Delta V_\perp \approx$  0.3mV, 0.8mV, 1.1mV respectively at $\approx$0.5V 3T background bias. This smoothness in the IETS from devices with the strongest MR signal $\Delta V_\perp$ is consistent with a merging of the energy levels of localized states distributed throughout the tunnel barrier, shifting their energetic position due to electric field, as illustrated in the insets of Fig. \ref{FIG2}. Thinner barriers (i.e. less resistive) have stronger internal electric fields under bias, leading to sharper, more energetically separated IETS peaks but lower absolute MR signals due to fewer contributing secondary inelastic scattering pathways; on the other hand, thicker barriers have more embedded impurities, leading to higher MR signals but a denser energetic distribution at a given voltage, which washes out the individual contributions from broadened IETS peaks. Others have explored the possibility that the 3T voltage signal arises from Coulomb repulsion and exchange effects in impurity tunneling states.\cite{Ando_arxiv2014,Song_arxiv2014}

\begin{figure}[t!]
\begin{center}
\includegraphics[width=8.5cm,height=7cm]{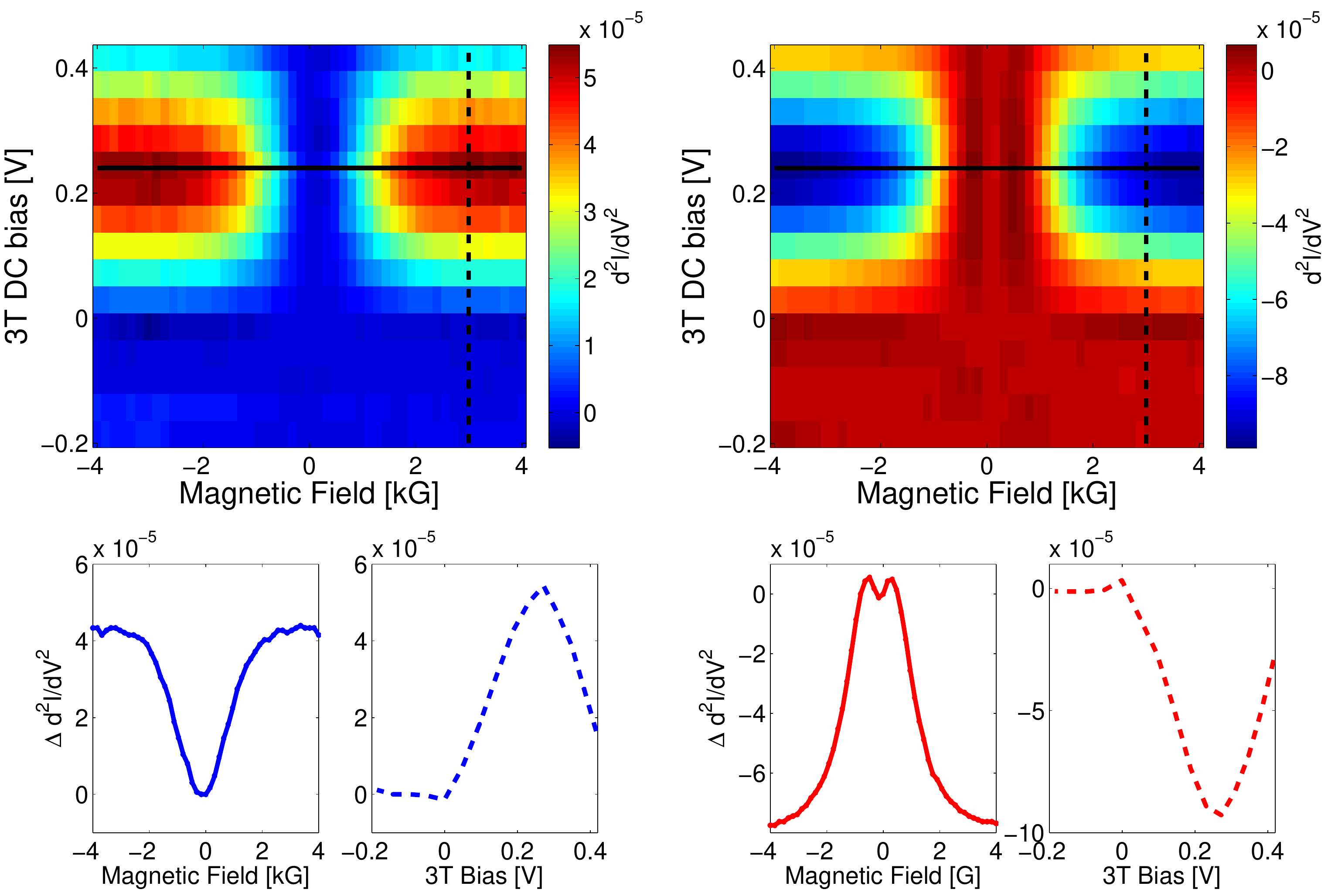}
\caption{(Color online) $\Delta \frac {d^2I}{dV^2}$ ($\frac {d^2 I(B)}{dV^2}-\frac {d^2 I(B=0)}{dV^2}$) as a function of magnetic field and dc bias taken at room temperature, $V_{AC}$=50mV, $f$=931Hz. The left (right) figures are data taken under perpendicular (parallel) field with specific cuts indicated by the solid and dashed lines.\label{FIG3}} 
\vspace{-25pt}
\end{center}
\end{figure}

We also measured the IET signal as a function of magnetic field at fixed voltage bias, shown in Fig. \ref{FIG3}. For the magnetic field oriented perpendicular and parallel, the IETS magnetic profile appears to be an inversion of the voltage signal shown in Fig. \ref{FIG1}(c). This profile grows and subsequently reduces at positive biases but is not present under negative bias. The overall change in second derivative as a function of field can easily be compared to $\Delta V_\perp$ through a Taylor power-law expansion; to lowest order approximation, $\Delta V_\perp \approx  \frac{1}{2}\Delta\left[\frac{d^2I}{dV^2}\right] V^2 R$, where $R$ is the differential resistance of the junction at the bias voltage in question. The maximum $\Delta \left[\frac{d^2I}{dV^2}\right]$ from the IETS in Fig. \ref{FIG3} corresponds to $\Delta V_\perp \approx$ 0.55 mV, which compares well to the measured value at the same background bias ($\approx$0.24V) of 0.7 mV. The fact that magnetic-field-induced changes in the IET spectra account for nearly all of the changes in the voltage signal indicates that the origin of the IETS signal (with contributions from inelastic tunneling) is closely linked to the origin of $\Delta V_\perp$.

\begin{figure}[t!]
\begin{center}
\includegraphics[width=8cm,height=3.75cm]{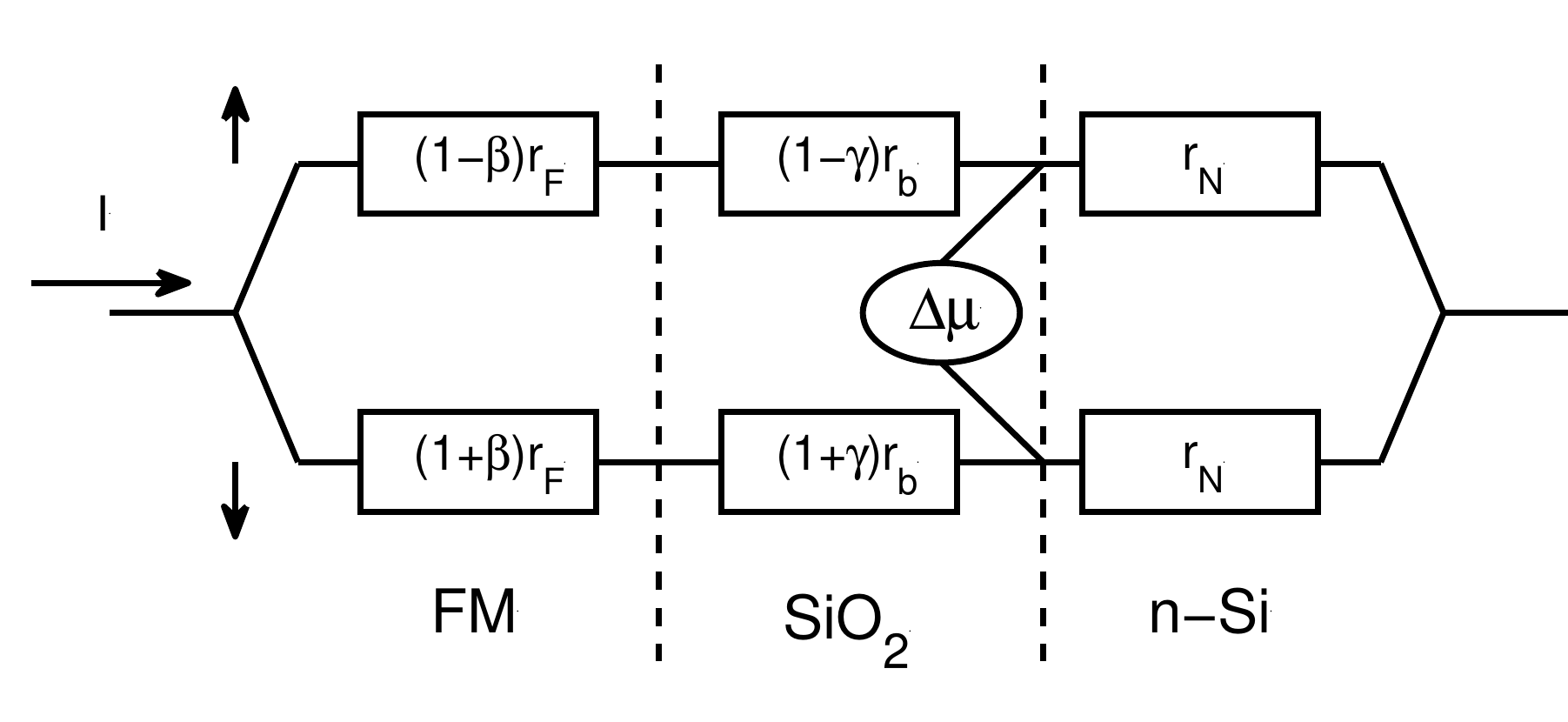}
\caption{(Color online) Resistor network model for spin majority (up) and minority (down) electrons flowing through the junction at a sourced current, showing relevant points where the electrochemical potential difference expressed in Eq. \ref{eq:t_B} is sensed.\label{FIG4}} 
\vspace{-25pt}
\end{center}
\end{figure}

We can provide additional quantitative evidence for this assertion with a simple two-current resistor network model of the tunnel junction, shown in Fig. \ref{FIG4}. The measured $V_\perp$ is related to the change in chemical potential $\Delta \mu$ by the expression $q \Delta V = \frac{\gamma}{2} \Delta \mu$, with $\Delta \mu$ given by 

\begin{equation}
\Delta \mu = q \frac{r_N(\beta r_F + \gamma r_b)}{r_N + r_F + r_b} I \; \approx q \gamma r_N I \;\; ( r_b \gg r_F,r_N)
\label{eq:t_B}
\end{equation}

\noindent where $r_N$, $r_F$, $r_b$ are the resistances of the nonmagnetic material, ferromagnetic material, and the barrier between them, respectively; the tunnel spin asymmetry $\gamma$ and ferromagnet bulk spin asymmetry $\beta$ coefficients are of order 0.1\cite{FertIEEE2007}. The nonlinear tunnel barrier resistance can be Taylor expanded to first order $r_b \approx r_{b0} + \frac{r_{b1}}{I_0} (I-I_0)$. Keeping terms up to quadratic order in $I$ and linear in $r_N$/$r_b$, we find the following expressions for the ratios of $\Delta V$ and its derivatives:
\begin{gather*}
\frac {\Delta V}{V} = \frac{\gamma^2 r_N}{2 r_{b0}} ,\nonumber\\
\frac {\Delta \frac {dV}{dI}}{\frac {dV}{dI}} = \frac{\gamma^2 r_N}{2 (r_{b0} - r_{b1})} \sim \frac{\Delta V}{V} ,\nonumber\\
\frac {\Delta \frac {d^2 V}{dI^2}}{\frac {d^2 V}{dI^2}} = - \frac{\gamma^2 r_N}{2 r_{b0}} \frac{(1+\beta / \gamma) r_F + r_N - 2 r_{b1}}{r_{b0}} \sim (\frac{\Delta V}{V})^2.\nonumber
\end{gather*}

\begin{figure}[t!]
\begin{center}
\includegraphics[width=8.5cm, height=7.5cm]{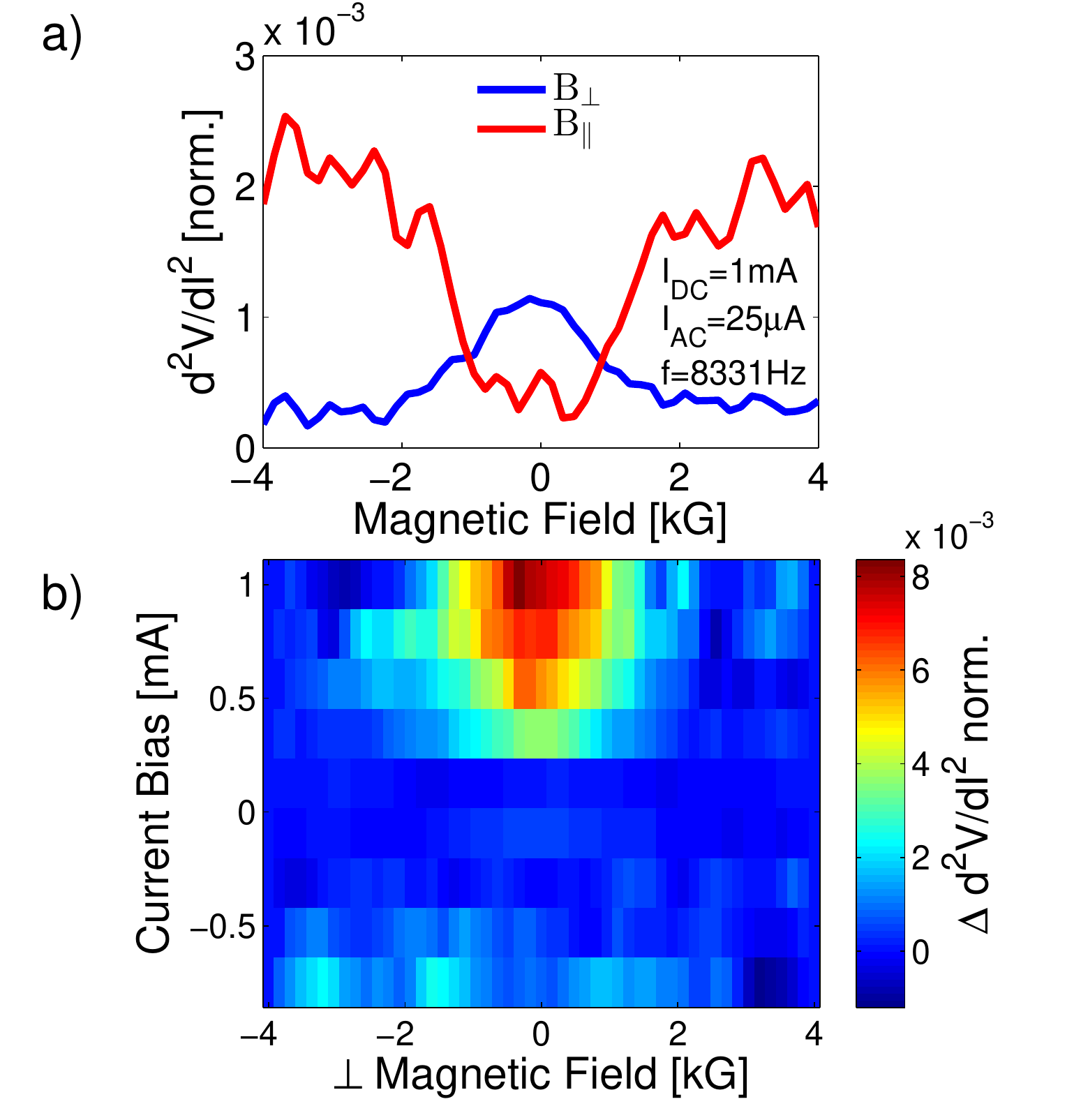}
\caption{(Color online) (a) RT measurements of $\frac {d^2 V}{dI^2}$ under perpendicular and parallel magnetic field at 1mA current bias normalized to the zero-field value. The $\frac {d^2 V}{dI^2}$ magnitude decreases significantly at higher currents, so the signal-to-noise ratio is much larger than the $\frac {d^2 I}{dV^2}$ measurements. (b) $\frac {d^2 V}{dI^2}$ under perpendicular field at various DC currents. Signals only appear for positive currents, consistent with previous measurements. \label{FIG5}} 
\vspace{-20pt}
\end{center}
\end{figure}

For these samples, $\frac {\Delta V}{V} \sim \frac{1{\text mV}}{1{\text V}} \sim 10^{-3}$, so we expect the corresponding ratio $\frac {\Delta \frac {d^2 V}{dI^2}}{\frac {d^2 V}{dI^2}}$ to be approximately $10^{-6}$ for spin accumulation signals. The IETS method only measures the second derivative in current, which can be mathematically related to the second and first derivative in voltage, but it is much simpler to directly measure the values instead. Using a similar lock-in technique in which an AC and DC current are applied to the junction and the second harmonic voltage response is detected, we measure changes in $\frac {d^2 V}{dI^2}$ orders of magnitude higher ($10^{-3}$, shown in Fig. \ref{FIG5}). This provides additional evidence, consistent with our lowest-order  analysis of $d^2I/dV^2$, that the large $\Delta V_\perp$ measured in the 3T scheme is not due to the Hanle effect or spin accumulation in the semiconductor conduction band. 

In conclusion, we have fabricated CoFe/SiO$_2$/n-Si junctions and measured magnetoresistance effects under parallel and perpendicular magnetic field in the 3T geometry in which the same contact is used to inject current and detect resulting changes in voltage. We find that $\Delta V_\perp$ does not follow an expected linear relationship with the applied current and is essentially absent in electron extraction. Because the detection contact is under large bias, the measured voltage is highly susceptible to secondary inelastic scattering events within the barrier, and we have employed an IETS technique to investigate the contribution of these inelastic conduction channels to the total current. The IETS shows a strong magnetic field dependence unaccounted for by simple spin accumulation into bulk Si, asserting that inelastic scattering in the barrier is responsible for the observed magnetoresistance signal.

\begin{acknowledgments}
We acknowledge the support of the Maryland NanoCenter and its FabLab. This work was supported at U. of Maryland by the Office of Naval Research under contract N000141110637, the National Science Foundation under contracts ECCS-0901941 and ECCS-1231855, and the Defense Threat Reduction Agency under contract HDTRA1-13-1-0013. This material is based upon work supported by the National Science Foundation Graduate Research Fellowship under Grant No. DGE1322106. 
\end{acknowledgments}

%

\end{document}